\documentclass[12pt]{iopart}

\usepackage{epsf}

\usepackage{graphicx}
\usepackage{bm}
\begin{document}


\title{Absorption Imaging and Spectroscopy of Ultracold Neutral Plasmas }


\author{ T. C. Killian, Y. C. Chen,   P. Gupta,  S. Laha,  Y. N. Martinez,
P. G. Mickelson, S. B. Nagel, A. D. Saenz, and C. E. Simien}
\address{ Rice University, Department of Physics and Astronomy
and Rice Quantum Institute, Houston, Texas, 77005}

\date{\today}

\begin{abstract}
Absorption imaging and spectroscopy can probe the dynamics of an
ultracold neutral plasma during the first few microseconds after
its creation. Quantitative analysis of the data, however, is
complicated by the inhomogeneous density distribution, expansion
of the plasma, and possible lack of global thermal equilibrium for
the ions. In this article we describe methods for addressing these
issues. Using simple assumptions about the underlying temperature
distribution and ion motion, the Doppler-broadened absorption
spectrum obtained from plasma images can be related to the average
temperature in the plasma.

\end{abstract}


\maketitle
\section{Introduction}
Ultracold neutral plasmas \cite{kkb99}, formed by photoionizing
laser-cooled atoms near the ionization threshold, stretch the
boundaries of traditional neutral plasma physics. The electron
temperature in these plasmas is from 1-1000K and the ion
temperature is around 1 K. The density can be as high as $10^{10}$
cm$^{-3}$. Fundamental interest in these systems stems from the
possibility of creating strongly-coupled plasmas \cite{ich82}, but
collective modes \cite{kkb00}, recombination \cite{klk01},  and
thermalization \cite{scg04} have also been studied.

Charged particle detection techniques have traditionally been used
for these experiments. However, optical absorption imaging and
spectroscopy, demonstrated \cite{scg04} using the Sr$^+$
${^2S_{1/2}} \rightarrow {^2P_{1/2}}$ transition in a strontium
plasma, opens many new possibilities. Images depict the density
profile of the plasma, and the Doppler-broadened absorption
spectrum measures the ion velocity distribution. Both can probe
 ion dynamics with 50\,ns resolution.

Qualitative interpretation of the images and spectrum is
straightforward, but quantitative analysis is complicated by the
inhomogeneous density distribution of the plasma, plasma
expansion, and the possible lack of global thermal equilibrium for
the ions.
 In order to address these effects, we are forced to make
 some simple assumptions about the form
of the ion temperature distribution and expansion of the plasma.
In this article we motivate these assumptions and describe our
methods for quantitatively analyzing the data.


Section \ref{Overview} provides an overview of the plasma creation
and absorption imaging technique. Section \ref{plasmadynamics}
describes the dynamics of the ions during the first few
microseconds after photoionization. This provides the basis  for
understanding Sec.\ \ref{spectrumsection}, which explains how the
absorption spectrum is extracted from the images and how it
relates to the temperature distribution.

\section{Experimental Overview}
\label{Overview}

\begin{figure}[t]
    \includegraphics[width=3in,clip=true,trim=55 220 20 90,angle=0 ]{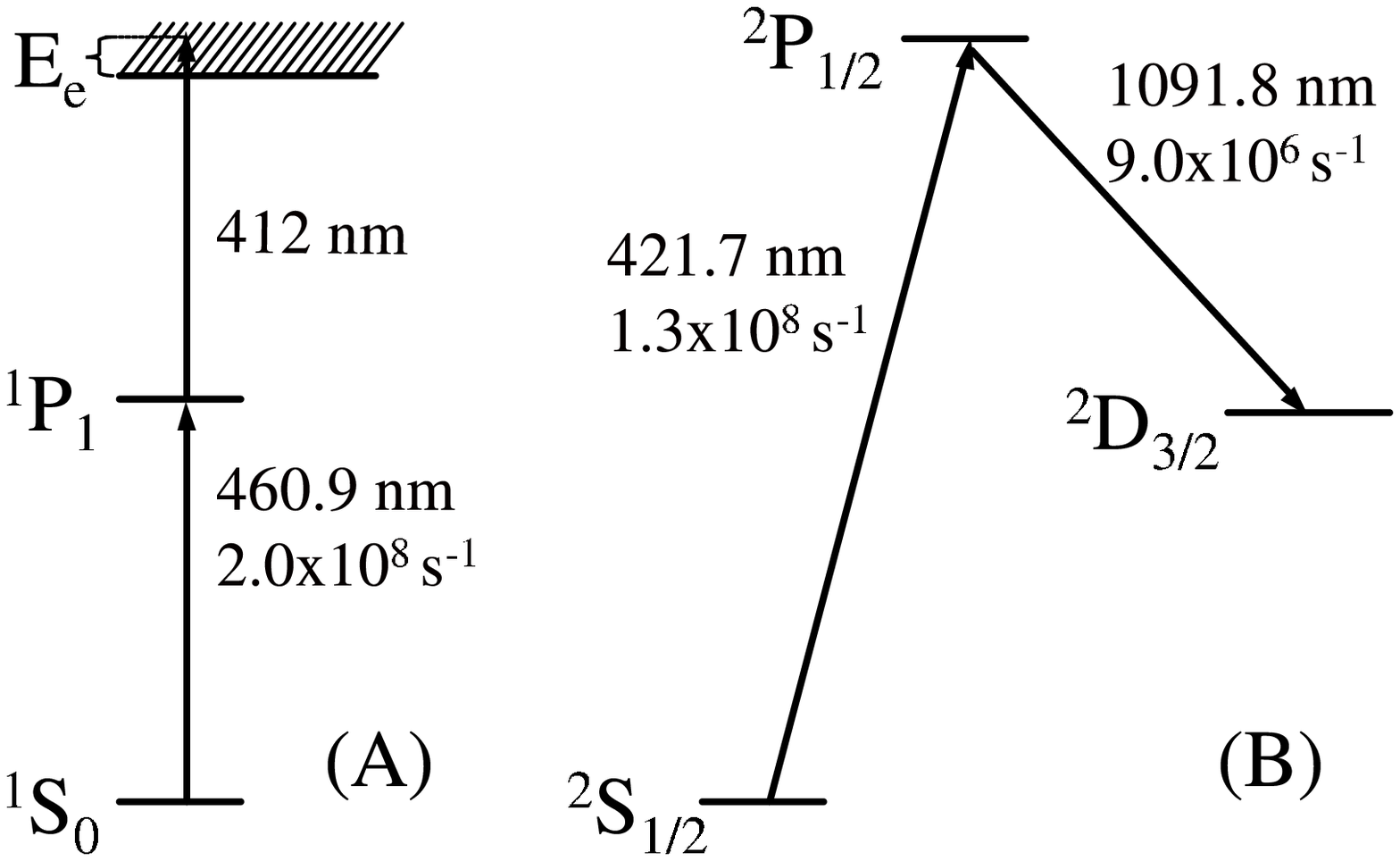}
  \caption{ Strontium atomic and ionic energy levels involved in the experiment, with decay rates.
  (A) Neutral atoms are laser cooled and trapped in a magneto-optical trap (MOT) operating on the
   ${^1S_0}-{^1P_1}$ transition at 460.9 nm, as
described in \cite{nsl03}. Atoms  excited to the $^1P_1$ level by
the MOT lasers are ionized by photons from a laser at $\sim
412$~nm.
 (B) Ions are imaged using the $^2S_{1/2}-{^2P_{1/2}}$ transition at $421.7$~nm.
 $^2P_{1/2}$ ions decay to the $^2D_{3/2}$ state 7\% of the time, after which
 they cease to interact with the imaging beam.
This does not complicate the experiment because ions typically
scatter fewer than one photon during the time the imaging beam is
on.}\label{energylevels}
\end{figure}

\begin{figure}
\includegraphics[width=4in,clip=true,trim=55 370 20 160,angle=0 ]{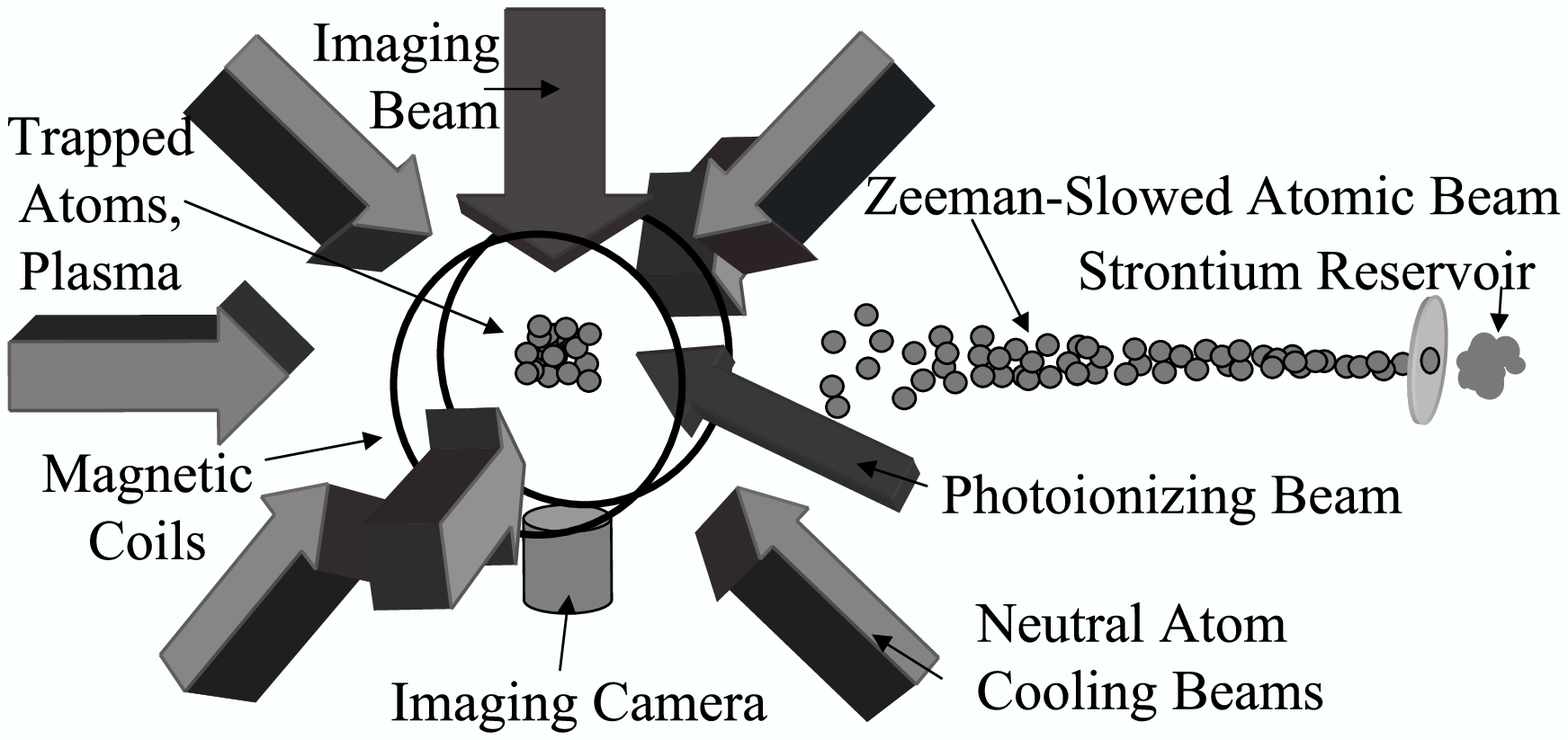}\\
  \caption{From \cite{scg04}.
  Experimental schematic for strontium plasma experiments.
   The MOT for neutral atoms consists of a pair of anti-Helmholtz magnetic
  coils and 6  laser-cooling beams. Atoms from a Zeeman-slowed
  atomic beam enter the MOT region and are trapped.
  {$^1P_1$} atoms are ionized by the photoionizing laser.
  The imaging beam passes through the plasma and
falls on a CCD camera.}\label{apparatus}
\end{figure}

The production of a strontium plasma starts with atoms that are
cooled and confined in a magneto-optical trap (MOT) (Figs.\
\ref{energylevels} and \ref{apparatus}). This aspect of the
experiment was described in \cite{nsl03}. The neutral atom cloud
is characterized by a temperature of a few mK and a density
distribution given by $n({r})=n_0{\rm exp}(-r^2/2\sigma^2)$, with
$\sigma \approx 0.6$~mm and $n_0 \approx 6 \times
10^{10}$\,cm$^{-3}$. The number of trapped atoms is typically $2
\times 10^8$. These parameters can be adjusted. In particular,
turning off the trap and allowing the cloud to expand yields
larger samples with lower densities.

To form the plasma, the MOT magnets are turned off and atoms are
ionized with photons from the cooling laser and from a $10$~ns
pulsed dye laser whose wavelength is tuned just above the
ionization continuum (Fig.\ \ref{energylevels}). Up to $30$\% of
the neutral atoms are ionized, producing plasmas with a peak
electron and ion density as high as $n_{0e}\approx n_{0i}\approx 2
\times 10^{10}$\,cm$^{-3}$. The density profiles, $n_{e}(r)\approx
n_{i}(r)$, follow the Gaussian shape of the neutral atom cloud.

Because of the small electron-ion mass ratio, the electrons have
an initial kinetic energy approximately equal to the difference
between the photon energy and the ionization potential, typically
between 1 and $1000$\,K. The initial kinetic energy for the ions
is close to the kinetic energy of neutral atoms in the MOT. As we
will discuss below, the resulting non-equilibrium plasma evolves
rapidly.

\begin{center}
\begin{figure}[t]
  \includegraphics[width=3in,clip=true]{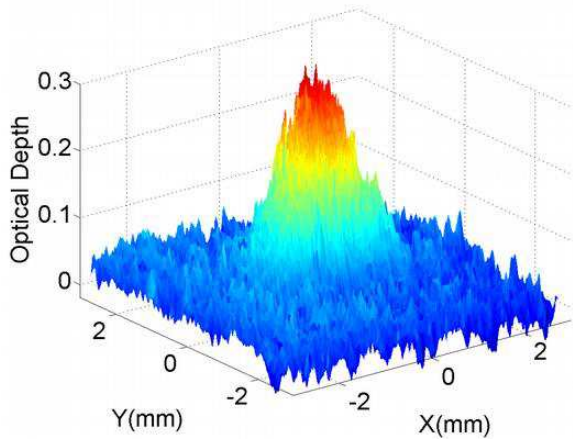}\\
  \caption{Optical depth of an ultracold neutral plasma.
  The delay between the formation of the plasma and
image exposure is $85$~ns.  The plasma contains
  $7 \times 10^7$ ions and the initial 
  peak ion density is $n_{0i}=2 \times 10^{10}$~cm$^{-3}$.
  Resolution is about $65$\,$\mu$m, limited
  by pixel averaging to improve the signal-to-noise ratio.
  }\label{image}
\end{figure}
\end{center}

To record an absorption image of the plasma, a collimated laser
beam, tuned near resonance with the principle transition in the
ions, illuminates the plasma and falls on an image intensified CCD
camera. The ions scatter photons out of the laser beam and create
a shadow that is recorded by an intensified CCD camera. The
optical depth ($OD$) is defined in terms of the image intensity
without ($I_{background}$) and with ($I_{plasma}$) the plasma
present,
\begin{eqnarray}\label{ODexperiment}
OD(x,y)&=&{\rm ln}(I_{background}(x,y)/I_{plasma}(x,y)) .
\end{eqnarray}
 Figure \ref{image} shows a typical absorption
image. Section \ref{spectrumsection} describes how detailed
information about the plasma is extracted from the optical depth.

\section{Ion Dynamics}
\label{plasmadynamics} In order to understand the details of the
image analysis, it is necessary to understand the dynamics of the
plasma.
 The imaging probe is most sensitive to the ion dynamics, so we will concentrate
 on this topic. The behavior of electrons was studied experimentally in
 \cite{kkb99, kkb00, klk01} and theoretically in
 \cite{kon02, mck02, rha03}.

 Ions are created with very little kinetic energy, but
 their initial spatially uncorrelated state
possesses significant Coulomb potential energy
  compared to the regular lattice that represents the
ground state of the system \cite{mbd98, mbh99}. As ions
equilibrate and correlations develop, the kinetic energy
increases. This process is called disorder-induced heating, or
correlation heating, and it has been discussed in many theoretical
papers. Early interest was generated by non-equilibrium plasmas
created by fast-pulse laser irradiation of solid targets, for
example \cite{bsk97, zwi99, mbm01, mno03}, and the problem has
been revisited in the context of ultracold neutral plasmas
\cite{kon02, mck02, mur01, ppr04jphysb}.

Qualitatively, one expects the ion temperature after equilibration
to be on the order of the Coulomb interaction energy between
neighboring ions. A quantitative analysis \cite{mur01}, assuming
complete initial disorder and incorporating the screening effects
of the
 electrons, predicts an ion temperature of
\begin{eqnarray}\label{iontemp}
  T_i&=&{2 \over 3} {e^2 \over
 4\pi \varepsilon_0 a k_B}\mid \tilde{U} +{\kappa \over
 2}\mid .
\end{eqnarray}
 Here, $\kappa =a/\lambda_D$ where
 $\lambda_D=(\varepsilon_0 k_B T_e/ n_{e} e^2)^{1/2}$ is the Debye
 length. The
 quantity $\tilde{U}\equiv {U \over N_i e^2 /
 4\pi \varepsilon_0 a} $ is the excess potential energy per particle
 in units of $e^2/4\pi \varepsilon_0 a$,
 where  $a=(4\pi n_i/3)^{-1/3}$ is the Wigner-Seitz radius, or interparticle
distance. $N_i$ is the number of  ions. $\tilde{U}$ has
 been studied with molecular dynamics simulations \cite{fha94}
 for a homogeneous system of particles interacting through a
 Yukawa
 potential, $\phi(r)= {e^2 \over 4\pi \varepsilon_0 r}{\rm
 exp}(-r/\lambda_D)$, which describes ions in the background of weakly
 coupled electrons
 \footnote{As the number of electrons per Debye sphere ($\kappa^{-3}$)
approaches unity, the Yukawa interaction ceases to accurately
describe ion-ion interactions. For strontium plasmas studied here,
this situation only occurs for the highest $n_e$ and lowest $T_e$.
It will be interesting to test Eq.\ \ref{iontemp} for these
conditions.}.

 For typical strontium plasmas discussed here,
 $\kappa \approx 0.1 - 1$, and $\lambda_D \approx 2 - 8$\,$\mu$m.
 $\tilde{U}$
 ranges from $-0.6$ to $-0.8$, so $T_i$ is close to
 ${e^2/
 4\pi \varepsilon_0 a k_B}$ as expected. $\kappa$ is related to the
Coulomb coupling parameter for electrons, $\Gamma_e$, through
  $\kappa=\sqrt{3\Gamma_e}$. A system is strongly coupled when
  $\Gamma>1$ \cite{ich82}.
  $\Gamma_e\approx 0.1 - 0.5$ for the systems studied here, so the
  electrons are not strongly coupled. This
  avoids excessive complications that arise when $\Gamma_e$
approaches or initially exceeds unity, such as screening of the
ion interaction \cite{kon02}, and rapid collisional recombination
and heating of the electrons \cite{kon02,mck02,rha03,tya00},
although we do see some signs of these effects, even in this
regime. The ions typically equilibrate with $T_i\approx1$\,K,
which gives $\Gamma_i\approx 3$, so the ions are strongly coupled.

 The time scale for disorder-induced heating is the inverse of the
 ion plasma
 oscillation frequency,
 $1 /\omega_{pi}= \sqrt{m_i \varepsilon_0/n_{i}
e^2}$, which is on the order of 100 nanoseconds. Physically, this
is the time for an ion to move about an interparticle spacing when
accelerated by a typical Coulomb force of $ {e^2 /
 4\pi \varepsilon_0 a^2}$. This time scale is also evident in
 molecular dynamics
 simulations  of ion-ion
 thermalization \cite{kon02, mck02, bsk97, zwi99, mbm01, mno03, mur01,
ppr04jphysb}.

It is interesting to note that under usual conditions in weakly
interacting plasmas or even atomic gases, the time scale for
relaxation of  the two-particle distribution function, which
describes correlations, is much faster than the collision time
that governs the relaxation of the one-particle distribution
function to the Maxwell-Boltzmann form. This is known as
Bogoliubov's hypothesis \cite{nic92}. For strongly-coupled
plasmas, however, these time scales both become equal to the
inverse of the plasma oscillation frequency \cite{zwi99}.

As the two-particle distribution function equilibrates, the
kinetic energy of the ions  exhibits strongly damped oscillations
at twice the ion plasma
 oscillation frequency. Intuitively, this can be understood as the
 oscillation of each ion in its local potential energy well. It is
 questionable whether this should be called an ion plasma oscillation
 or not because there is probably no collective
 or long range coherence to
 the motion. This behavior has been observed in molecular dynamics
 simulations of equilibrating strongly-coupled systems
 \cite{zwi99,mno03,ppr04jphysb}. The damping time for the
 oscillations is approximately $ \pi / \omega_{pi}$ for $\Gamma \ge
 5$. Averaging over the entire density distribution, as we do in the analysis
 described here, obscures the oscillations. A different approach,
 which resolves regions of different density in the plasma, is
 required to clearly observe the oscillations, and this phenomenon
 will not be discussed further in this paper.




For $t_{delay}> \pi /\omega_{pi}$, the ions have equilibrated and
the thermal energy of the electrons begins to dominate the
evolution of the plasma. Electrons contained in the potential
created by the ions exert a  pressure on the ions that causes an
outward radial acceleration. This was studied experimentally in
\cite{kkb00} and theoretically by a variety of means in
\cite{rha03}. The experiments measured the final velocity that the
ions acquired, which was approximately $v_{terminal}\approx
\sqrt{E_e/m_i}$. With the imaging probe, we now observe the
expansion dynamics at much earlier times during the acceleration
phase.

As discussed in  \cite{kkb00} and \cite{scg04}, a  hydrodynamic
model, which describes the plasma on length scales larger than
$\lambda_D$, shows that the pressure of the electron gas drives
the expansion through an average force per ion of
\begin{eqnarray}\label{ionforce}
\bar{F}&=& {-{\bar \nabla}(n_e(r) k_B T_e) \over n_i(r)} \approx
\hat{r} {r k_B T_e \over \sigma_i^2},
\end{eqnarray}
 where the ion and electron density distributions are
 $n_{e}(r)\approx n_{i}(r)=n_{0i}{\rm
 exp}(-r^2/2\sigma_i^2)$. We
 assume thermal equilibrium for the electrons throughout the
 cloud \cite{rha03}.

 The force leads to an average radial expansion velocity for
 the ions,
 \begin{equation}\label{ionvelocity}
    \bar{v}(r,t_{delay})=\hat{r}{r k_B T_e \over m_i\sigma_i^2}t_{delay}.
\end{equation}
 The velocity is correlated
with position and increases linearly with time.
 This does not represent an increase in the random thermal
 velocity spread or temperature of the ions.
 Due to the large mass difference, thermalization of ions and
 electrons \cite{kon02} is slow and occurs on a millisecond time scale.



Equation \ref{ionvelocity} for the average ion velocity assumes a
constant electron temperature. Actually, as the plasma expands,
electrons will cool. This can be thought of in terms of energy
conservation or adiabatic expansion. It is possible to describe
the expansion  with a Vlasov equation  that includes the changing
electron temperature. For an initial Gaussian density
distribution, the equations can be solved analytically and the
expansion preserves the Gaussian shape with a $1/\sqrt{e}$ density
radius given by $\sigma_i^2(t) \approx \sigma_i^2(0) +[k_B
T_e(0)/m_i]t^2$ \cite{ rha03, dse98, ppr04}. The experiments
involving absorption imaging of the plasma, however, have
concentrated on the first few microseconds of the expansion when
the plasma size and electron temperature have not changed
significantly. Thus we can safely use Eq.\ \ref{ionvelocity}. The
effects of the expansion are evident in the radial velocity that
manifests itself in Doppler broadening of the ion absorption
spectrum.






\section{Doppler-Broadened Spectrum}
\label{spectrumsection}









\begin{figure}
  \includegraphics[width=3in,clip=true]{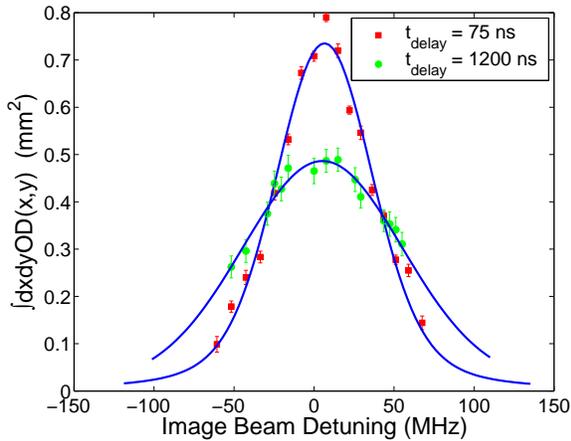}\\
  \caption{Absorption spectra of ultracold neutral plasmas.
  We plot the integral of the optical depth (Eq.\ \ref{ODintegral}).
  The frequency is with respect to a Doppler-free absorption
  feature in a strontium discharge cell.
  Both spectra correspond to $T_e=56$~K and
  an initial peak plasma density of
  $n_{0i}=2 \times 10^{10}$~cm$^{-3}$.
  Data are fit with Voigt profiles, and the increase in
  linewidth for longer $t_{delay}$ is clear.}\label{spectrum}
\end{figure}

To  obtain quantitative information from the plasma images, we
relate the $OD$ (Eq.\ \ref{ODexperiment}) to underlying physical
parameters. Following Beer's law, the $OD$ for a laser propagating
along the z axis is
\begin{eqnarray}\label{ODtheory}
OD(x,y)&=&\int dz \hspace{.025in}
       n_i(x,y,z) \alpha[\nu, T_i(r)],
\end{eqnarray}
 where
 $n_{i}(x,y,z)$ is the ion density, and $\alpha[\nu, T_i(r)]$ is the
 ion absorption cross section at
 the image beam
frequency, $\nu$. The  absorption cross section is a function of
temperature due to Doppler broadening, and since we expect the
temperature to vary with density, we allow $\alpha$ to vary with
position. If we now integrate over x and y, or, in reality, sum
over the image pixels multiplied by the pixel area, we get the
spectrum
\begin{eqnarray}\label{ODintegral}
S(\nu)\equiv \int dx dy OD(x,y)&=&\int d^3r \hspace{.025in}
       n_i(r) \alpha[\nu, T_i(r)],
\end{eqnarray}
as a function of the image laser detuning \footnote{We can also
fit $OD(x,y)$ to a two dimensional Gaussian, as described in
\cite{scg04}, and identify  $\int dx dy OD(x,y)\approx 2\pi
\sigma_{ix}\sigma_{iy} OD_{max}$, where $\sigma_{ix}$ and
$\sigma_{iy}$ are the transverse sizes of the absorption profile,
and $OD_{max}$ is the peak optical depth. This sometimes has
signal-to-noise ratio advantages over integrating the entire
image, but both approaches should give the same result.}.  As we
vary the detuning, we obtain absorption spectra as shown in Fig.\
\ref{spectrum}. The rest of the paper will deal with the
relationship between spectra such as these and the underlying
temperature distributions of the ions.

 The
absorption cross section for ions in a region described by a
temperature $T_i$, is given by the Voigt profile
\begin{eqnarray} \label{absorptioncrossection}
  \alpha(\nu, T_i)&=&\int d s {3^*\pi \lambda^2 \over
  2}{1 \over 1+ 4( { \nu-s \over \gamma_{eff}/2\pi} )^2} {1
  \over \sqrt{2\pi} \sigma_D(T_i)} {\rm e}^{-(s-\nu_0)^2/2\sigma_D(T_i)^2},
\end{eqnarray}
where $\sigma_D(T_i)=\sqrt{k_B T_{i}/m_i}/\lambda$ is the Doppler
width, and $\gamma_{eff}=\gamma_0+\gamma_{laser}$ is the effective
Lorentizian linewidth due to the natural linewidth of the
transition, $\gamma_0=2\pi \times 22 \times 10^6$\, rad/s, and the
laser linewidth, $\gamma_{laser}=2\pi \times (10 \pm 2)\times
10^6$\, rad/s. The center frequency of the transition is
$\nu_0=c/\lambda$, where $\lambda=422$\,nm. The ``three-star"
symbol, $3^*=1$, accounts for the equal distribution of ions in
the doubly degenerate ground state and the linear polarization of
the imaging light \cite{sie86}.

Extracting the ion temperature from the spectrum is complicated by
the fact that  we do not expect to have global thermal equilibrium
of the ions during the first microsecond after plasma formation.
Global thermal equilibrium will occur on a hydrodynamic time
scale, $\sigma_i/v$, which is on the order of ten microseconds,
where $v$ is the ion acoustic wave velocity \footnote {In
principle, it is possible to obtain spectra from small regions of
the cloud, so as to examine the local dynamics and avoid averaging
over the plasma. Developing this capability will be the subject of
future studies.}.

We need a  method to relate the Doppler broadening of the spectrum
to a characteristic temperature.
We will show that for reasonable models of the ion temperature
distribution, the temperature extracted from a fit of a Voigt
profile to the spectrum yields a temperature that is a close
approximation of the average ion temperature in the plasma. More
thorough checks of this assignment, such as with molecular
dynamics simulations, would be valuable, but this serves as a
useful working definition for discussing the data.

\begin{figure}
  \includegraphics[width=3in,clip=true]{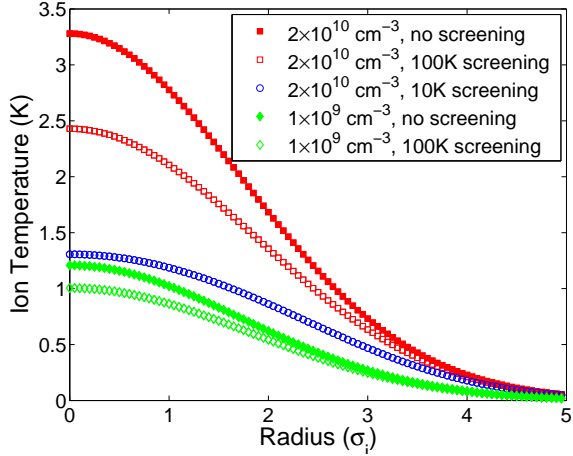}\\
  \caption{Ion temperature profiles. Peak densities ($n_{0,i}\approx
  n_{0,e}$) and electron temperatures are given in the legend.
Screening refers to the iterative calculation of temperature using
Eq.\ \ref{iontemp}. No screening refers to Eq.\
\ref{localfinaliontemperature}, which neglects electron screening,
and is thus independent of electron temperature. In the limit of
low density and high electron temperature, both expressions give
the same result. }\label{CheckTformula}
\end{figure}

For times longer than $\pi/\omega_{pi}$, but short compared to
hydrodynamic times, we expect local thermal equilibrium at a
temperature approximately given by Eq.\ \ref{iontemp}. Under this
model, the temperature  varies across the plasma because the
interparticle spacing, $a$, varies with density. Given  $T_e$ and
$n_i$, it is possible to use Eq.\ \ref{iontemp} in an iterative
recipe to find the local ion temperature. This procedure utilizes
an expression \cite{hfd97} for
the excess potential energy $\tilde{U}$ in terms of $\Gamma_i$ and
$\kappa$ that is valid for  $\kappa<5$. This yields ion
temperature distributions as shown in Fig.\ \ref{CheckTformula}.


To simplify the discussion, we first neglect electron screening.
This accurately describes the plasma for $\kappa \ll 1$  (high
$T_e$ and low $n_e$). To a good approximation, we can then say
\begin{eqnarray}
  T_i(r)&=&C{e^2 \over
 4\pi \varepsilon_0 a(r) k_B}=T_{i,max}{\rm
 e}^{-r^2/6\sigma_i^2},\label{localfinaliontemperature}
\end{eqnarray}
where $T_{i,max}$ is the temperature corresponding to the peak
density at $r=0$. Molecular dynamics simulations of ultracold
plasmas that neglect electron screening display this thermal
distribution  when the ions have equilibrated locally
($t_{delay}$\raisebox{-.6ex}{$\stackrel{>}{\sim}$} $\pi /
\omega_{pi}$) \cite{ppr04private}. $C=0.45$ is a constant
determined by numerical evaluation of the parameters in Eq.\
\ref{iontemp} for small $\kappa$ \footnote{As $\kappa$ increases,
Eq.\ \ref{localfinaliontemperature} still describes the ion
temperature if $C$ decreases and becomes a function of position in
the plasma, as shown in Fig.\ \ref{CheckTformula}.}.
The average ion temperature, assuming
Eq.\ \ref{localfinaliontemperature}, is
\begin{eqnarray} \label{averagetemperature}
  T_{i,ave}={1 \over N_i}\int d^3r \hspace{.025in}
       n_i(r) T_i(r)=T_{i,max}{3\sqrt{3} \over 8}.
\end{eqnarray}

\begin{figure}
  \includegraphics[width=4in]{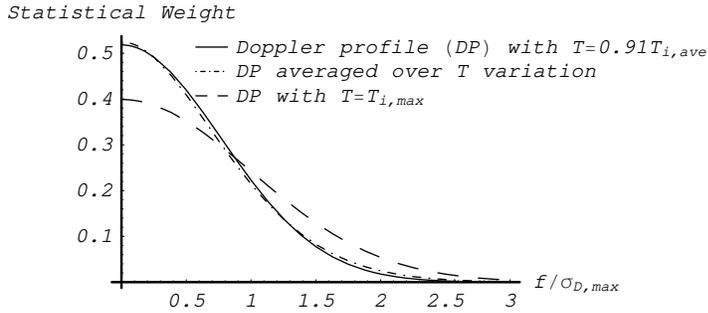}\\
  \caption{
  Assuming $T_i(r)$
  is given by Eq.\ \ref{localfinaliontemperature},
  the Doppler profile averaged over the density distribution
  can be replaced by a
 Doppler profile characterized by a single uniform temperature,
 $T=0.59 \hspace{.025in} T_{i,max} =0.91\hspace{.025in}T_{i,ave}$. The statistical weight is
 proportional to the probability of finding ions with that Doppler shift, and the
 horizontal axis is the Doppler shift in units of $\sigma_{D,max}$.}
  \label{averagedoppler}
\end{figure}

 Combining the expressions
for the integral over the optical depth (Eq.\ \ref{ODintegral}),
and the expression for the absorption cross section (Eq.\
\ref{absorptioncrossection}) yields
\begin{eqnarray}\label{bigODintegral1}
S(\nu)&=&\int d^3r \hspace{.025in}
       n_i(r) \int d s {3^*\pi \lambda^2 \over
  2}{1 \over 1+ 4( { \nu-s \over \gamma_{eff}/2\pi} )^2} 
{1 \over \sqrt{2\pi} \sigma_D[T_i(r)]} {\rm
  e}^{-(s-\nu_0)^2/2\sigma_D[T_i(r)]^2}.
\end{eqnarray}
If we insert  (Eq.\ \ref{localfinaliontemperature}) for $T_i(r)$,
we find
\begin{eqnarray}\label{bigODintegral2}
S(\nu) = \int \hspace{.025in} d s {3^*\pi \lambda^2 \over
  2}{1 \over 1+ 4( { \nu-s \over \gamma_{eff}/2\pi} )^2} \int d^3r \hspace{.025in}
      n_{0i}{\rm e}^{-r^2/2\sigma_i^2} \times  \nonumber \\
       {1
  \over \sqrt{2\pi} \sigma_{D,max}}{\rm e}^{r^2/12\sigma_i^2} {\rm
  exp}[-{\rm e}^{r^2/6\sigma_i^2}(s-\nu_0)^2/2\sigma_{D,max}^2].
\end{eqnarray}
Here $\sigma_{D,max}$ is the Doppler width corresponding to
$T_{i,max}$. As shown in Fig.\ \ref{averagedoppler}, the average
over the density distribution of the Doppler factor can be
replaced by a single Doppler distribution with temperature $T=0.59
\hspace{.025in}T_{i,max} = 0.91 \hspace{.025in}T_{i,ave}$.
\begin{eqnarray}\label{averagedopplereq}
 \int d^3r \hspace{.025in} n_{0i}
      {\rm e}^{-r^2/2\sigma_i^2}  \nonumber {1 \over \sqrt{2\pi} \sigma_{D,max}} {\rm e}^{r^2/12\sigma_i^2}
{\rm
  exp}[-{\rm e}^{r^2/6\sigma_i^2}(s-\nu_0)^2/2\sigma_{D,max}^2] \nonumber \\
   \approx {N_i \over \sqrt{2\pi} \tilde{\sigma}_{D}}{\rm
  exp}[-(s-\nu_0)^2/2\tilde{\sigma}^{2}_{D}],
\end{eqnarray}
where $\tilde{\sigma}_{D}=\sqrt{0.91 \hspace{.025in} k_B
T_{i,ave}/m_i}/\lambda$. So if the velocity distribution of the
ions is characterized by local thermal equilibrium of the form in
Eq.\ \ref{localfinaliontemperature},  fitting the spectrum,
$S(\nu)$, to a Voigt profile yields a good approximate measure of
the quantity $T_{i,ave}$.

If we want to include the effects of electron screening, we can
follow exactly the same procedure, except in our model of the
spectrum (Eq.\ \ref{bigODintegral1})  we use the iterative recipe
for ion temperature using Eq.\ \ref{iontemp}, instead of  the
simple Eq.\ \ref{localfinaliontemperature}. This slightly more
complicated numerical analysis shows that if Eq.\ \ref{iontemp}
describes the temperature distribution, then the temperature
extracted from the Voigt profile is betwen $0.91$ and $0.95$ of
$T_{i,ave}$. In this case, the average ion temperature is found by
averaging Eq.\ \ref{iontemp} over the density distribution.

If we have global thermal equilibrium, which we expect for long
$t_{delay}$, then the temperature is uniform, and, by definition,
the Voigt profile will yield the average temperature.

The expansion of the plasma, Eq.\ \ref{ionvelocity}, also
contributes to the velocity distribution. So we must expand our
model. Fortunately, the expansion is well understood
\cite{kkb00,rha03,ppr04} and determined by the parameters $T_e$
and $\sigma_i$. Independent analysis of the images determines
$\sigma_i$. $T_e$ is approximately equal to $2 E_e/3k_B$, where
$E_e$ is the detuning of the photoionizing laser above resonance
\cite{kkb99}, although for high density and low electron
temperature, we can expect a small increase of $T_e$ above $2
E_e/3k_B$ due to electron heating from three-body recombination
\cite{rha03}, disorder-induced heating \cite{kon02}, and continuum
lowering \cite{mck02}.

 We will first treat the expansion using the simple model of ion
 temperature, Eq.\ \ref{localfinaliontemperature}.
 The expansion velocity profile  gives rise to an average Doppler shift
of the resonant frequency that varies with position,
\begin{eqnarray}\label{doppler shift}
 \delta \nu=v_z/\lambda&=&{r  k_B T_e t_{delay} \over m_i \sigma_i^2 \lambda}cos\theta \nonumber\\
&=&\sigma_{D,ave} {r\over \sigma_i}  {t_{delay} \over
t_{exp}}cos\theta,
\end{eqnarray}
where $rcos\theta$ is  the displacement from the center of the
cloud along the  direction of laser propagation. We have
introduced a characteristic time for the expansion,
\begin{eqnarray}\label{texpansion}
t_{exp}=&\sigma_i  \sqrt{ m_i \over k_B T_e} \sqrt{T_{i,ave} \over
T_e},
\end{eqnarray}
 which
is the time at which the Doppler shift due to expansion, at
$rcos\theta=\sigma_i$,  equals $\sigma_{D,ave}$. In other words,
this is the time at which the Doppler broadening due to expansion
becomes comparable to the thermal Doppler broadening. For typical
plasma conditions, such as $T_e=30$\,K, $T_{i,ave}=1$\,K, and
$\sigma_i=1$\,mm, Eq.\ \ref{texpansion} yields
$t_{exp}=3.4$\,$\mu$s.

\begin{figure}
  \includegraphics[width=2.25in, angle=270]{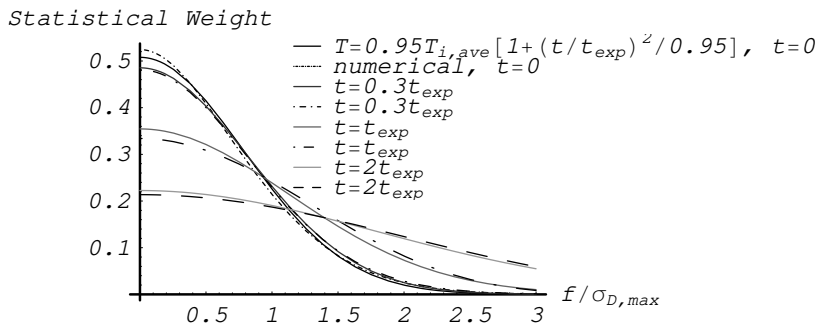}\\
  \caption{Doppler profile averaged over the
  density distribution as in Fig.\ \ref{averagedoppler},
but now including the effects of expansion.
The dashed or dash-dot numerical expressions are full integrals
over volume shown in Eq.\ \ref{expansiondoppler}. The solid curves
are the Gaussians for Doppler broadening arising from a single
temperature $T = 0.95 \hspace{.025in}
T_{i,ave}[1+(1/0.95)(t/t_{exp})^2]$, where $t_{exp}$ is the
characteristic time scale for the expansion described in the text.
At all times, the single Doppler profile is a good
approximation to the full integral.  
}
  \label{averagedopplerwithexpansion}
\end{figure}

We include the position-dependent Doppler shift in the exponent of
the Gaussian describing the  Doppler broadening in the Voigt
convolution (Eq.\ \ref{bigODintegral2}), which becomes
\begin{eqnarray}\label{expansiondoppler}
 \int d^3r \hspace{.025in}n_{0i}
      {\rm e}^{-r^2/2\sigma_i^2} {1 \over \sqrt{2\pi} \sigma_{D,max}} {\rm e}^{-r^2/12\sigma_i^2}
    {\rm exp}[-{\rm e}^{r^2/6\sigma_i^2}({s-\nu_0
}-\sigma_{D,ave}{r\over \sigma_i} {t_{delay} \over
t_{exp}}cos\theta)^2/2\sigma_{D,max}^2] \nonumber \\
   \approx {N_i \over \sqrt{2\pi } \sigma_{D}(T_{i,eff})} 
   {\rm
  exp}[-{(s-\nu_0)^2 \over
  2\sigma_{D}^2(T_{i,eff})}].
\end{eqnarray}
In the second line of Eq.\ \ref{expansiondoppler}, we have given
the Doppler profile, corresponding to a temperature of
\begin{eqnarray} \label{effectivetemperature}
  T_{i,eff}&=& C \hspace{0.025in}
  T_{i,ave}[1+{1 \over C}({t_{delay} \over t_{exp}})^2],
\end{eqnarray}
 that
approximately equals the original full integral, including
variation in temperature across the plasma and effects of
expansion. Here, $C=(0.95 \pm0.05)$. Figure
\ref{averagedopplerwithexpansion} shows that the agreement is
good over the entire time scale of the expansion.

If we include screening effects and use Eq.\ \ref{iontemp} to find
the temperature, then numerical simulation shows that the
extracted temperature is also given by Eq.\
\ref{effectivetemperature}, with $C$ ranging from $0.91$ to
$0.95$. If the plasma has reached global thermal equilibrium, then
it can be shown analytically that $T =
T_{i,ave}[1+(t_{delay}/t_{exp})^2]$ exactly. This suggests  that
when fitting  a Voigt profile to the spectrum derived from the
integral over the optical depth,
\begin{eqnarray}\label{voigtalleffects}
S(\nu)&=& N_i \int d s {3^*\pi \lambda^2 \over
  2}{1 \over 1+ 4( { \nu-s \over \gamma_{eff}/2\pi} )^2}
{1
  \over \sqrt{2\pi} \sigma_D(T_{i,eff})} {\rm
  e}^{-(s-\nu_0)^2/2\sigma_D(T_{i,eff})^2},
\end{eqnarray}
the extracted effective temperature can be related to the average
temperature in the plasma through Eq.\ \ref{effectivetemperature}.
One can show analytically that Eq.\ \ref{effectivetemperature} has
the correct form in the limit when all broadening is negligible
except the Doppler broadening due to expansion.

Our motivation of Eq.\ \ref{effectivetemperature} relies on
certain models of the ion temperature. They are reasonable
approximations to the distributions seen in molecular dynamics
simulations except at short times, $t<\pi /\omega_{pi}\approx
500$\,ns, during the initial rapid disorder-induced heating and
oscillatory phase when  the kinetic energy distribution as a
function of temperature is not well known. Because this time scale
varies with density, we expect different regions to equilibrate
and oscillate at different times. Extracted ion temperatures
should be viewed as more qualitative for these early times.

\begin{figure}
  \includegraphics[width=5in,clip=true]{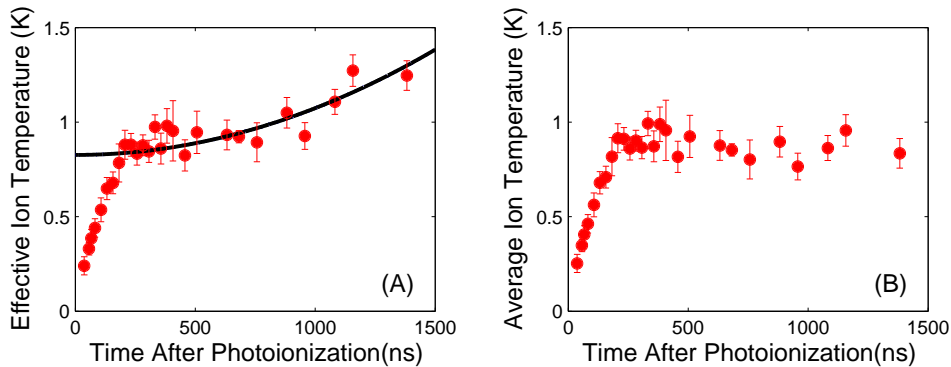}
  \caption{Evolution of the ion temperature in an ultracold neutral plasma.
  The initial 
  peak ion density is $n_{0i}=5 \times 10^{9}$~cm$^{-3}$ and $\sigma_i=780$\,$\mu$m. The plasma contains
  $6 \times 10^7$ ions.
   (A) The effective ion temperature corresponding to the Doppler
   broadening extracted from Voigt profile fits of data such as
   in Fig.\ \ref{spectrum}. The solid line
   (Eq.\ \ref{effectivetemperature})
   is the effective
   temperature one would expect for $T_{i,ave}=0.9$\,K and
   $T_e=2 E_e/3k_B=40$\,K.
   Equation \ref{effectivetemperature} includes the effects of expansion but neglects early
   equilibration.
  (B) The average ion temperature is obtained by dividing by
  $0.95 \hspace{0.025in}[1+(1/0.95)(t_{delay}/t_{exp})^2]$. 
  }\label{tempevolutiondata}
\end{figure}
Equation \ref{effectivetemperature} is the central result of this
 paper.
 It enables us to
 analyze the data in a clear and well-defined way, as shown in
 Fig.\ \ref{tempevolutiondata}. 
 It also
 separates the various contributions to the Doppler broadening of
 the spectrum.
For example, Fig.\ \ref{tempevolutiondata}A shows the time
dependence of $T_{i,eff}$ for a typical plasma. Figure
\ref{tempevolutiondata}B shows the extracted $T_{i,ave}$ with the
expansion effects removed. The only fit parameter in the model
line in Fig. \ref{tempevolutiondata}A  is $T_{i,ave}$. This
procedure allows us to determine $T_{i,ave}$ to an accuracy of
about 0.2\,K. Alternatively, after removing the effect of
expansion by scaling the data by $0.95
\hspace{0.025in}[1+(1/0.95)(t_{delay}/t_{exp})^2]$, as in
\ref{tempevolutiondata}B, the equilibration temperature can be
identified by the plateau. Scaling the data also improves the
visibility of the disorder-induced heating during the first $
1/\omega_{pi}\approx 180$\,ns. When there is significant electron
heating \cite{kon02,rha03,mck02}, $T_e$ will exceed  $3E_e/2k_B$,
and $T_e$ then becomes a fit parameter that provides information
on the electron dynamics.


Using this recipe for analyzing the data we can conduct
quantitative studies of the rate and equilibration temperature for
disorder-induced heating. We can also examine the factors
affecting ion temperature oscillations, electron temperature, and
plasma expansion. These will be the subjects of future work.

We thank A. Chan, T. Pohl, T. Pattard, and J. M. Rost for helpful
discussion. This research was supported by the Department of
Energy Office of Fusion Energy Sciences, National Science
Foundation, Office for Naval Research, Research Corporation,
Alfred P. Sloan Foundation, and David and Lucille Packard
Foundation.

\vspace{.25 in}


\end{document}